\documentclass[12pt]{iopart}
\usepackage{graphicx}  
\begin{document}
\title{Photon strength distributions in stable even-even molybdenum isotopes.}
\author{
A~Wagner$^1$,
R~Beyer$^1$,
M~Erhard$^1$,
E Grosse$^{1,2}$,
A R Junghans$^1$,
J Klug$^1$,
K Kosev$^1$,
C Nair$^1$,
N Nankov$^{1,3}$,
G Rusev$^1$,
K D Schilling$^1$,
R Schwengner$^1$
}
\address{
$^1$Institut f\"ur Strahlenphysik,
Forschungszentrum Dresden-Rossendorf, 
PF 510119, 01324 Dresden, Germany}
\address{
$^2$Institut f\"ur Kern- und Teilchenphysik,
Technische Universit\"at Dresden,
01062 Dresden, Germany}
\address{
$^3$Institute for Nuclear Research and Nuclear Energy,
Bulgarian Academy of Sciences,
1784 Sofia, Bulgaria}
\ead{a.wagner@fzd.de}
\begin{abstract}
Electromagnetic dipole-strength distributions up to the particle
separation energies are studied for the stable even-even nuclides 
$^{92,94,96,98,100}$Mo in photon scattering experiments at the 
superconducting electron accelerator ELBE of the Forschungszentrum 
Dresden-Rossendorf.
The influence of inelastic transitions to low-lying excited states 
has been corrected by a simulation of $\gamma$ cascades using 
a statistical model.
After corrections for branching ratios of ground-state transitions, 
the photon-scattering cross-sections smoothly connect
to data obtained from $(\gamma,n)$-reactions.
With the newly determined electromagnetic dipole response of nuclei 
well below the particle separation energies the parametrisation 
of the isovector giant-dipole resonance is done with improved
precision.
\end{abstract}
\submitto{\JPG}
\section{Introduction}
The most prominent excitation mode in nuclei is constituted by the 
Isovector Giant Dipole Resonance (GDR) which can be interpreted 
macroscopically as the oscillation of proton matter against neutron 
matter.
Nevertheless, the low-energy tail of the GDR strength at energies 
well below the maximum of the GDR has not been studied with 
high accuracy, so far.
Theoretical predictions of the low energy tail of the GDR vary between
pure Lorentzian distributions (constant width) and modified 
Lorentzian distributions (energy dependent width) \cite{Kadmenskii:1983}. 
Additionally, at excitation energies of about 6 MeV, recent photon
scattering experiments have shown extra strength commonly dubbed
``pygmy resonance''.
Since the accurate knowledge of the photoabsorption cross section 
close to the particle separation energies is of importance for the
modelling of astrophysical processes in hot scenarios like the
$\gamma$-process \cite{Goriely:2002} we started an experimental 
programm to investigate systematically the evolution of electromagnetic 
dipole strength in the stable even-even isotopes $^{92,94,96,98,100}$Mo.

\section{Photon-scattering experiments}
In principle, the photoabsorption cross sections $\sigma_\gamma$ can be 
measured via $\gamma$ rays emitted after photoexcitation. 
However, the increasing density of nuclear states with increasing excitation
energy results in de-excitation patterns which are combinations of 
direct ground-state decays and transitions via intermediate states.
Due to the vast number of transitions observed in the deexciation 
of medium-mass nuclei it reveals impossible to attach the observed
transitions to nuclear states.
Therefore, we developed an approach to unfold measured 
photon-scattering yields on the basis of a statistical model.
Above the particle separation energies, the photoabsorption cross
section has been determined previously in $(\gamma,n)$-reactions.

The experiments were carried out at the bremsstrahlung facility at the
superconducting electron accelerator ELBE of the Forschungszentrum
Dresden-Rossendorf. 
Bremsstrahlung was produced by electrons impinging onto a 3.4 mg/cm$^2$ 
thick Nb radiator.
A narrow photon beam is formed by an aluminium collimator with a length of 
2.6 m and a conical opening angle of 5 mrad while the electrons
are deflected by a purging magnet.
An absorber made of a 10 cm long Al cylinder between the radiator and the 
collimator attenuates the intense low energy part of the bremsstrahlung 
spectrum. 
The photon beam passed the target about 570 cm downstream inside an evacuated 
polyethylene beam pipe and was subsequently absorbed in a well shielded 
photon-beam dump. 
Scattered photons were registered in four high-purity Ge detectors (HPGe) 
with efficiencies of 100 \% relative to a 3"$\times$3" NaI detector. 
In order to determine the multipole order of the scattered $\gamma$ rays 
two of the detectors were located at $127^\circ$ and the
other two at $90^\circ$ with respect to the photon beam at distances of 32 and
28 cm to the target, respectively. 
The low-energy photons are suppressed by lead absorbers with thicknesses
0.8 and 1.3 cm at the two given angles, respectively, combined with 0.3 cm 
thick copper absorbers. 
The HPGe detectors are equipped with 3 cm thick escape-suppression shields 
made of BGO scintillation detectors. 
The detector resolution amounts to 5.0 keV (7.9 keV) at 5 MeV (9 MeV)
photon energy.
The bremsstrahlung facility of the ELBE accelerator is
described in detail in Refs. \cite{Schwengner:2005, Wagner:2005}. 

\begin{figure}
\begin{center}
\includegraphics[width=0.7\textwidth]{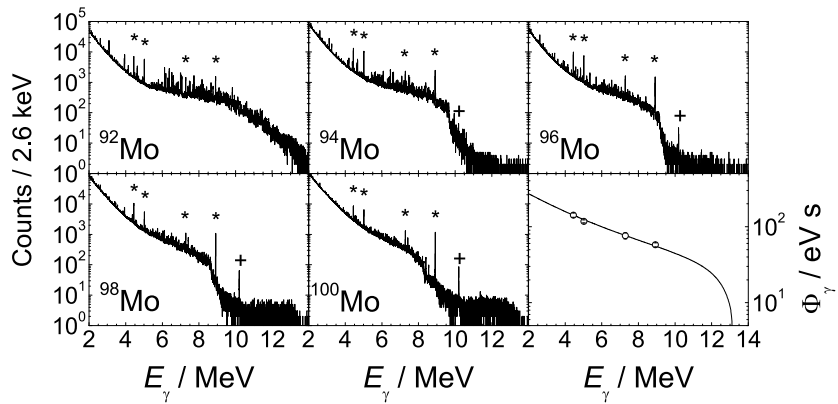}
\caption{\label{fig:1}Experimental spectra of $\gamma$-rays scattered from
$^{92,94,96,98,100}$Mo and the distribution of the incident bremsstrahlung
produced by an electron energy of 13.2 MeV. The spectra are taken at an 
angle of 127$^\circ$ relative to the photon beam.
Transitions marked by stars belong to the calibration standard $^{11}$B,
while crosses mark $\gamma$ rays from $^{73}$Ge$(n,\gamma)$ reactions.}
\end{center}
\end{figure}

We performed all measurements at identical experimental conditions and the 
same electron-beam energy $E_e^{\rm kin}=13.2$ MeV. 
Samples of elementary $^{92,94,96,98,100}$Mo isotopically enriched to 
more than
97 \% with masses of 
2036 mg, 1998 mg, 2003 mg, 2952 mg, and 2916 mg, respectively, 
were used as targets. 
The spectra of photons scattered at $127^\circ$ from 
$^{92,94,96,98,100}$Mo measured for 51 h, 105 h, 95 h, 59 h, and 57 h, 
respectively, are presented in Fig.~\ref{fig:1}.
The detector response function is obtained using simulations based on the
GEANT3 Monte-Carlo code and calibrated radiation standards.
The incident photon flux is determined by photon scattering from 
well-known transitions in $^{11}$B and is shown in Fig.~\ref{fig:1} as well.

\section{Determination of photon strength functions}
The electric dipole photon strength for photon absorption is defined as the 
average reduced transition width
\begin{equation}
\stackrel{_\rightarrow}{f}_{E1}(E_\gamma) = E_\gamma^{-3}
\frac{\langle \Gamma^{E1}_0(E_\gamma) \rangle}{D},
\end{equation}
with the average level width $\langle \Gamma^{E1}_0(E_\gamma)\rangle$ 
at energy $E_\gamma$ and the average level spacing $D$.
The dipole photon strength can be translated into the average
photon absorption cross section $\langle \sigma_\gamma\rangle$ with 
the ground state spin $J_0$ and the spin of the excited state $J$:
\begin{equation}
\stackrel{_\rightarrow}{f}_{E1}(E_\gamma) = \frac{2J_0+1}{2J+1}
\frac{\langle \sigma_\gamma(E_\gamma) \rangle}{(\pi \hbar c)^2E_\gamma}.
\end{equation}
The magnetic dipole strength can be determined accordingly. 
We obtained the $M1$ strengths in $^{92,98,100}$Mo from discrete transitions 
and calculated an average photon absorption cross section of about 1 mb
at excitation energies from 7 - 8 MeV \cite{Rusev:2006} which is the region
of spin-flip excitations.
The contribution of $M1$-strength is taken into account in the determination
of the dipole strength later on.

\begin{figure}
\begin{center}
\includegraphics[width=0.7\textwidth]{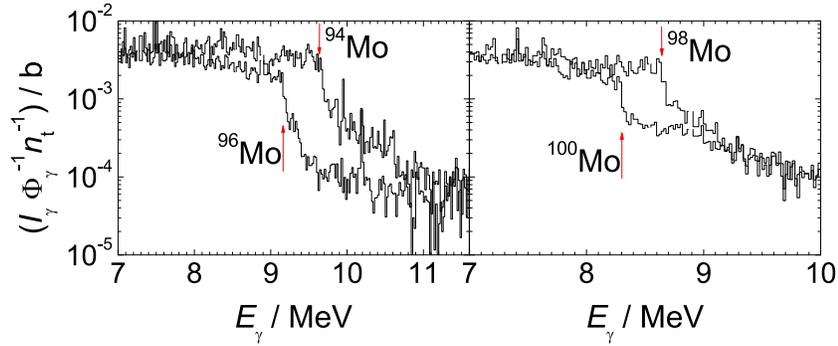}
\caption{\label{fig:2}Spectra of $\gamma$-rays scattered from
$^{94,96}$Mo (left) and $^{98,100}$Mo (right) around the neutron
separation energies (labelled by arrows) divided by the incident 
photon flux and the respective number of target atoms.}
\end{center}
\end{figure}

The intensity distribution obtained from the measured spectra after a
correction for detector response and a subtraction of atomic background in the
target contains a continuous part in addition to the resolved peaks which 
accounts for about 70\% of the total strength.
Fig.~\ref{fig:2} shows the spectra of scattered photons close to the 
neutron separation energies which demonstrates the significant amount 
of unresolved strength as compared to strength in resolved transitions.
In order to estimate the intensities of inelastic transitions to low-lying 
levels we have applied statistical methods 
\cite{Rusev:2007,Becvar:1998,Becvar:2007}. 
By means of simulations of $\gamma$-ray cascades intensities of
transitions to low-lying states could be removed from this intensity 
distribution and the intensities of ground-state transitions could be 
corrected for their branching ratios $\Gamma_0 / \Gamma$ leading to 
the determination of the photoabsorption cross section 
$\langle\sigma(E_\gamma)\rangle$.

\section{Parametrisation of the GDR for triaxially deformed nuclei}
The chain of stable molybdenum isotopes is a good example for a study of
the dipole strength distributions at the onset of (triaxial) deformation. 
The ground state deformations evolve
from a deformation parameter $\beta=0.04$ for the closed neutron-shell 
nucleus $^{92}$Mo via $\beta=0.05 (0.08,0.18)$ for the transitional
nuclei $^{94}$Mo ($^{96}$Mo, $^{98}$Mo) to $\beta=0.24$ 
for $^{100}$Mo~\cite{Moeller:1995}.
The experimentally deduced shapes of the considered Mo isotopes are soft where
the triaxiality parameter $\gamma$ for $^{98}$Mo is $32^\circ$ 
\cite{Zielinska:2002}.
Therefore, the properties of the investigated Mo isotopes are dominated by
quadrupole deformation and triaxial shape.
The influence of deformation on the E1 strength at the low energy tail
of the GDR has been studied microscopically for medium-mass nuclei 
in \cite{Donau:2007}.
 
According to the hydrodynamical model, the GDR is represented as a vibration of
the proton system against the neutron system. 
For nuclei with stable deformation in the ground state, the GDR splits into
independent vibrations along the principal axes of the nucleus 
conserving the total strength. 
The integrated photoabsorption cross section corresponding to vibrations
along one of the axes is given by the
Thomas-Reiche-Kuhn dipole-sum rule 
$S_{\rm TRK}=\int{\sigma_\gamma(E_x) dE_x=60 ZN/A}$ mb
MeV \cite{Dietrich:1989} while the resonance energies $E_i$ are inversely 
proportional to the length of the axis:
\begin{equation}
\label{eq:EiGDR}
E_i=E_0 \exp\bigg(-\sqrt{\frac{5}{4}\pi}\beta
\cos(\gamma-\frac{2}{3}\pi i) \bigg).
\end{equation}
$E_0$ is the energy of the maximum of the GDR of a spherical nucleus. 
The Hill-Wheeler parameters $\beta$ and $\gamma$ used in Eq. (\ref{eq:EiGDR}) 
can be taken from Ref. \cite{Rusev:2006}.

We attempt to parametrise the low-energy tail of the GDR by a 
sum of three Lorentzians. Considering the wide range of excitation energy 
spanned by the combined data, a test of a Lorentzian with energy-dependent 
total width $\Gamma(E_x)$ is indicated:
\begin{equation}
\sigma_\gamma(E_x) = \frac{2 C \cdot S_{\rm TRK}}{3 \pi}
\sum_{i=1}^{3}\frac{E^2_x \Gamma(E_x)}{(E_i^2 - E^2_x)^2 + E^2_x 
\Gamma(E_x)^2}.
\end{equation}
The parameter $C$ measures the conformance of the integrated $E1$ strength 
with the Thomas-Reiche-Kuhn sum-rule. We use the parametrisation 
$\Gamma(E_x)=\Gamma_S \cdot (E_x/E_i)^\delta$ of the energy dependence of the
width with $\delta$ as a parameter to be defined by a fit to the
combined data.

\begin{figure}
\begin{center}
\includegraphics[width=0.7\textwidth]{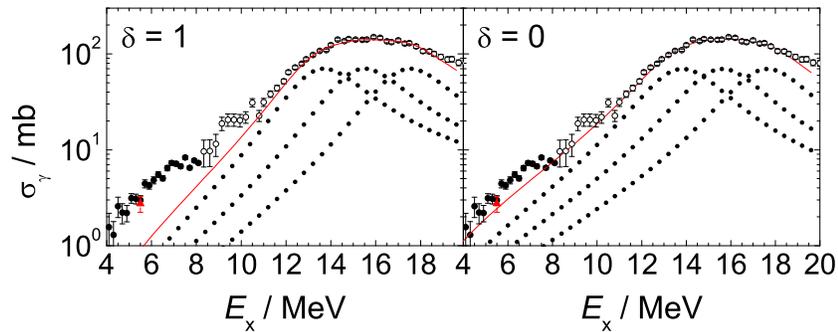}
\caption{\label{fig:3}Comparison of photoabsorption cross sections
for $^{100}$Mo determined in the photon scattering experiment and in
$(\gamma,n)$-reactions \cite{Beil:1974} with two different parametrisations 
of the Giant Dipole Resonance. 
The parametrisation consists of a superposition of three Lorentzians 
according to the triaxial deformation of $^{100}$Mo using a linear 
enery dependence of the width (left) and an energy-independent width
(right).
Data obtained in neutron capture experiments are shown as red (color online)
triangles.}
\end{center}
\end{figure}

\begin{figure}
\begin{center}
\includegraphics[width=0.7\textwidth]{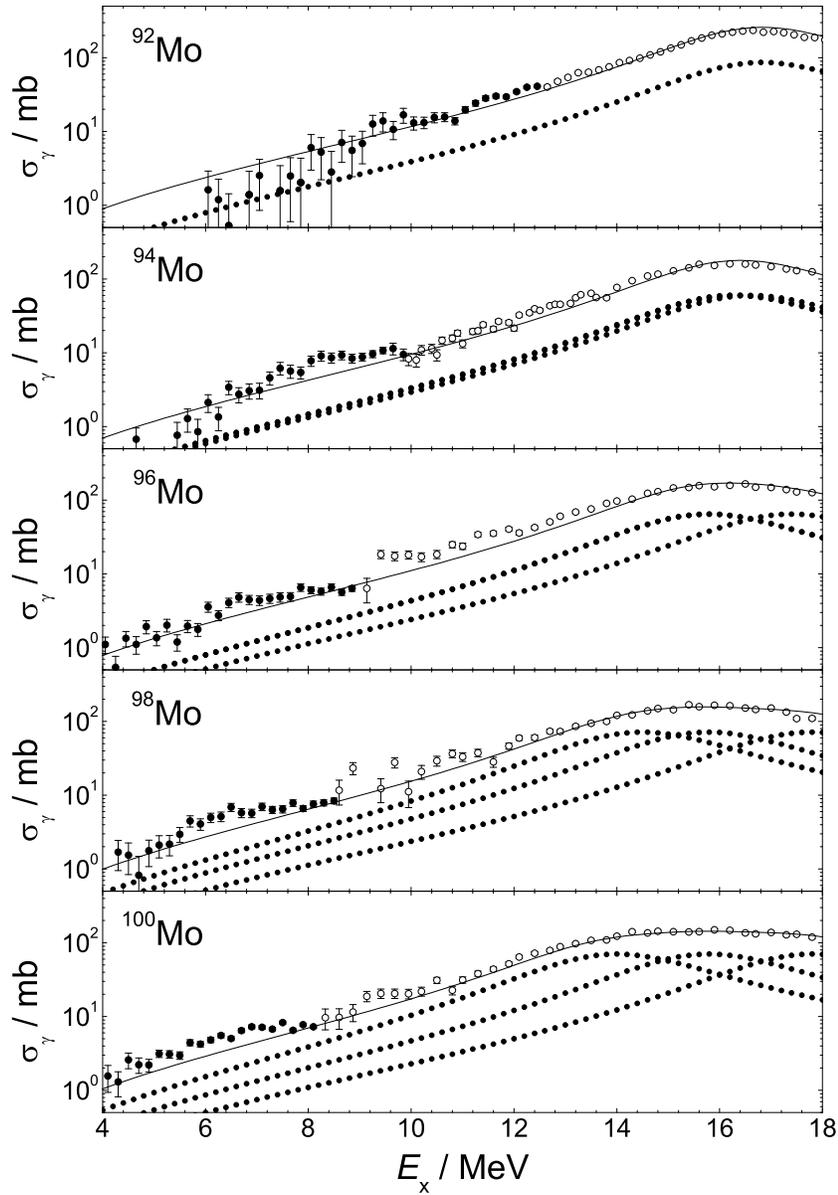}
\caption{\label{fig:4}Photoabsorption cross sections for all stable even-even
molybdenum isotopes derived from photon scattering (filled symbols) 
and $(\gamma,n)$-reactions (open symbols)\cite{Beil:1974}.
The solid lines indicate the parametrisation using the sum of up to 
three Lorentzians (dashed lines).}
\end{center}
\end{figure}

The coupling of the particle-hole states to more complex configurations
leads to an effective damping and consequently a wider Lorentzian shape.
Nevertheless, a long standing question on a proposed energy dependence of
the damping width \cite{Wambach:1988,Carlos:1974} was never finally 
answered due to the lack of precise and systematic data on the 
photon strength function. 
In this work, we obtain $\delta$ = 0.0(4), i.e., no energy dependence of the 
width within the uncertainty. 
Fig.~\ref{fig:3} shows two different parametrisations using a linear
energy dependence width as compared to a constant width.
In accordance to our findings in the neighbouring nucleus 
$^{88}$Sr \cite{Schwengner:2007} and to the systematics of the GDR widths 
\cite{Carlos:1974} we use $\Gamma_S=4$ MeV. 
For the parameter $C$ we find values of 
1.08(1), 0.77(1), 0.83(1), 0.92(2), 0.90(1) 
from the fit of the ($\gamma$,$n$) data of 
$^{92}$Mo, $^{94}$Mo, $^{96}$Mo, $^{98}$Mo, and $^{100}$Mo, 
respectively. 
The results of the parametrisation are presented in Fig.~\ref{fig:4}.

By the proposed clear-cut separation of the two causes for the widening of the
GDR (deformation and damping), we obtain an expression of use also when no
GDR-data are available, i.e. for nuclei far off stability
entering nucleosynthesis calculations. 
Without much influence on the off-resonance cross section the GDR-energy can 
then be taken from systematics
\cite{Bohr:1975}, and by using $C=1$ and calculated deformation parameters
\cite{Rusev:2006}, the dipole strength from 4-5 MeV up to threshold can be
predicted.

This work was supported by the Deutsche Forschungsgemeinschaft under contract
DO-466/1-2.

\end{document}